\documentstyle[sprocl]{article}

\def\Journal#1#2#3#4{{#1} {\bf #2}, #3 (#4)}

\def\be{\begin{equation}}
\def\ee{\end{equation}}
\def\bea{\begin{eqnarray}}
\def\eea{\end{eqnarray}}

\begin{document}

\title{VIBRONIC MECHANISM OF THE ISOTOPE AND PRESSURE EFFECTS IN CUPRATE}

\author{A. S. MOSKVIN, Y. D. PANOV}

\address{Department  of Theoretical Physics, Ural State University, \\
Lenin av.51, 620083, Ekaterinburg, Russia\\
E-mail: alexandr.moskvin@usu.ru}

\maketitle\abstracts{ The doped cuprates are considered to be a system of local spinless bosons moving in a lattice 
of the hole pseudo-Jahn-Teller centers $CuO_{4}^{5-}$. A  detailed qualitative and 
quantitative description of the vibronic mechanism determining both the isotope substitution 
and the external pressure effect on $T_c$  is given.  This mechanism can properly interpret 
the principal peculiarities of isotopic and baric effects in cuprates except the region of 
the well developed percolation phenomena}

Intuitive ideas concerning an especial role played by the Jahn-Teller (JT) ions (centers, polarons) 
have been used as a starting point of the pioneer investigations by K.A. M\"uller and J.G. 
Bednorz accomplished  in 1986 by the outstanding discovery of the high-$T_c$ superconductivity. 
One of the most realistic JT-effect based scenarios for doped cuprates like 
$La_{2-x}Sr_{x}CuO_4$ was suggested and elaborated earlier \cite{Moskvin}.
The most exciting features of the hole centers  $CuO_{4}^{5-}$  are 
associated with an unusually complicated ground state resulting from the electronic 
quasi-degeneracy accompanied by the (pseudo)-Jahn-Teller (PJT) effect. An additional hole 
doped to the basic $CuO_{4}^{6-}$ cluster with the $b_{1g}$ hole can occupy both the same 
hybrid $Cu3d-O2p$ orbital state resulting in a Zhang-Rice singlet $^1A_{1g}$ and the purely 
oxygen $e_u$ molecular orbital resulting in the singlet or triplet $^{1,3}E_u$ terms with the 
close energies. If the vibronic coupling is strong enough, the adiabatic potential (AP) of the hole 
$[CuO_4^{5-}]_{PJT}$ PJT center has the four  equivalent   minima, which 
correspond to the hybrid distortions of the $b_{1g}e_{u}$- or $b_{2g}e_{u}$-type, so the 
vibronic ground state is fourfold degenerated, if to neglect  a tunnel splitting.
The hole PJT center with its high polarizability could be a 
center of an effective local pairing with a formation of a novel electron PJT center 
$[CuO_{4}^{7-}]_{PJT}$. 

A novel superconducting phase in the doped cuprates appears to be a
result of the disproportionation reaction
\begin{equation}
2CuO_4^{6-}\rightarrow \left[ CuO_4^{5-}\right] _{JT}+\left[
CuO_4^{7-}\right] _{JT}, 
\end{equation}
with formation of system of the local spinless bosons moving in a lattice of the hole PJT 
centers  or the  quantum lattice bose liquid.
Bose-condensation temperature is proportional to the local boson transfer integral: 
$T_c = t_{BB}f(N_{B})$, where $f(N_{B})$ is a smooth function of the boson concentration 
\cite{Alexandrov}. 
The PJT nature of the polar (electron and hole) $CuO_{4}$ clusters provides a main vibronic 
mechanism both of the isotopic and baric effects on the $T_c$ due to  vibronic reduction 
of transfer integral \cite{MP} 
\begin{equation}
t_{BB}=t_{BB}^{(0)}\,\ K_{eh}
\end{equation}
with  the vibronic reduction factor $K_{eh}$, or the Franck-Condon factor
\begin{equation}
K_{eh}=\left\langle \Psi_{e}|\Psi_{h}\right\rangle ^{2}, \label{K}
\end{equation}
where $\Psi_{e}$ and $\Psi_{h}$ are the ground state vibronic functions
in a definite  AP minimum for electronic ($e$-) and hole ($h$-) 
center, respectively. 
These functions depend on different parameters  of the polar center 
(such as elastic and vibronic constants, crystal field parameters), 
so that any impact on the lattice or electron  subsystems can change the 
$K_{eh}$ magnitude. 
Real perspectives for a high-$T_{c}$ superconductivity should be connected
with an extremely weak inter-mode coupling regime or with the so called 
{\it optimized systems}. A boson movement in optimized systems is not 
accompanied by a significant variation in the spin and the local structure
modes or in other words the charge fluctuations do not result in the spin
and structure fluctuations. Electron and hole PJT centers in optimized
systems have identical vibronic states with the maximal value of the
"vibronic reduction" factor $K_{eh}=1$, so the vibronic contribution to the $T_c$ 
variation (suppression or enhancement) for these systems due to the isotope substitution, 
the external pressure application etc.,  is entirely supressed. In practice, 
the  specific features of the optimized
systems are the maximally high $T_{c}$'s, the minimal width of the
superconducting transition, the minimal magnitude of the isotope effect, the
minimal value of the baric coefficient $dT_{c}/dp$. Probably, the so called
"optimally doped" $YBa_{2}Cu_{3}O_{6+x}$ oxide at $x\approx 0.93$ can be
approximately considered as one of the real optimized systems. 
A detailed analysis of the  PJT effect within the $(^{1}A_{1g},^{1,3}E_u)$ manifold with 
account of the whole set of the active displacement modes of the
$a_{1g}$, $b_{1g} $, $b_{2g}$, 
$e_{u}$ symmetry allows to obtain  an analytical form for the reduction factor $K_{eh}$. 
This can be reduced to universal form: 
\begin{equation}
K_{eh}=D\ exp(-g^2), 
\end{equation}
where $D$ and $g$ in a complicated manner depend on the parameters like bare energy 
separations within $(^{1}A_{1g},^{1,3}E_u)$-manifold, vibronic constants, the oxygen and 
copper atomic masses. 
Detailed analytical examination of general expressions for $g$ and $D$ with account of 
real multi-mode situation is very complicated. 
But a case of isotope substitution appears to be essentially simplified because of only vibrational 
part of $K_{eh}$ plays a role.
For a simplest one-dimensional single-mode situation 
\begin{equation}
K_{eh}=\frac{2\tau }{1+\tau ^{2}}\exp \left( -\frac{\left( \Delta Q\right)
^{2}}{l_{e}^{2}+l_{h}^{2}}\right) , 
\end{equation}
where $l_{e}$ and $l_{h}$ are the effective oscillatory lengths of the $e$-
and $h$-centers, respectively, $\tau =l_{e}/l_{h}$, $g^{2}\sim $ $(\Delta Q)^{2}$, with 
$\Delta Q$ being the distance 
separating the AP minima for the $e$- and $h$-centers. For  two-dimensional
oscillators centered for simplicity at the same point $g=0$, thus $K_{eh}\equiv D$ and 
\[
K_{eh}=4\left[ \tau +\frac{1}{\tau }+ \alpha \cos ^{2}\Delta \phi + \beta  \sin ^{2}\Delta \phi
\right] ^{-1},
\]
\begin{equation}
 \alpha = \left( \frac{\tau _{e}}{\tau _{h}}+%
\frac{\tau _{h}}{\tau _{e}}\right),
\beta = \left( \tau_{e}\tau _{h}+\frac{1}{\tau _{e}\tau _{h}}\right), 
\end{equation}
where $\tau _{i}=l_{i}^{(1)}/l_{i}^{(2)}$ ($i=e,h$), $l_{i}^{(1,2)}$ -
oscillatory lengths, which are the principal axes of the polarization ellipse defined by
quadratic form for AP expansion near potential minima  for $e$- or $h$-center, respectively; 
$\tau =S_{e}/S_{h}$, $S_{i}$ is an area of ellipse; $\Delta \phi $ is  an angle between 
principal axes of ellipses for $e$- or $h$-center.

So, in a multi-mode case the $g$ contribution plays the same role, but $D$ factor 
provides two contributions to vibronic reduction, due to
variation in the oscillatory lengths ("$stretching$" mode) and due to variation in 
relative orientation of the polarization ellipses ("$tilting$" mode)  for the $e$- 
or $h$-center, respectively.

For isotope effect (IE) this leads to the following 
results. In the simplest one-mode case $l_{e}$
and $l_{h}$ have an identical dependences on mass: $l_{e,h}\sim m^{-\frac{1%
}{4}}$ , and thus $D$ factor doesn't depend on atomic masses.
Another factor $g^{2}\sim $ $\sqrt{m}(\Delta Q)^{2}$, so that $\alpha \geq 0$
and IE will be determined mainly by the distance $\Delta Q$
separating the AP minima for the PJT center in the $e$- and $h$%
-state that results in a considerable positive isotope coefficient 
(IC) at low $T_c$.
In a multi-mode case the $g$ factor contribution plays the same role, but the $D$
factor contribution can be both positive and negative.
The principal results of analytical consideration and computer simulation for the 
vibronic IE on the local boson transfer integral and $T_c$ in the system of the  PJT 
centers could be summarized as follows:

1. There is no simple $\alpha (T_{c})$ correlation, nonetheless the absolute value of the IC usually  increases with decreasing $T_{c}$. 
Minimal IE should be observed for the optimized systems with maximal $T_c$'s. However, this doesn't imply 
a negligible role of the electron-vibrational coupling. Within our model an extremely strong electron-vibrational 
(vibronic) coupling easily coexists with the negligibly small IE.

2. As a whole, the vibronic contribution to
IE appears to be relatively small, so even at favorable  conditions the
oxygen exponent $\alpha _{O}$ reaches the corresponding BCS value 
$\alpha_{BCS}=0.5$ only for small $T_{c}\sim 0.2\ T_{c}^{max}$. Anomalously large values $\alpha _{O}>1.0$ 
are possible only at $T_{c}<0.1\ T_{c}^{max}$. 

3. As a rule, both oxygen and copper IC appear to be positive. Negative IE could be observed  as a result 
of peculiar compensation of the positive $g$- and negative $D$-contributions. In practice, such a compensation 
could be realized only for copper due to occurrence of the only hybrid $e_u$ modes  and within a rather 
narrow range of the appropriate parameters and with a rather small numerical value $|\alpha _{Cu}|\leq 0.2$. 
Such a compensation is unlikely simultaneously for several oxygen modes. Quite  the contrary, negative copper 
IE $\alpha _{Cu}\leq 0$ is usually accompanied   by relatively large positive $\alpha _{O}$ values, while positive 
copper IE $\alpha _{Cu}\geq 0$ does by  moderate positive $\alpha _{O}$ values.

4. Numerical values of the copper and oxygen IC's are determined by the material-dependent quantities such as bare parameters $%
\Delta _{e},\Delta _{h}$ for the $^{1}A_{1g}-^{1,3}E_u$ separation, elastic and vibronic constants. Moreover, for instance, 
the most popular HTSC systems $YBa_{2}Cu_{3}O_{6+x}$ and  $La_{2-x}Sr_{x}CuO_{4}$ are distinguished by the  ground state 
adiabatic potential ($b_{1g}e_{u}$- and $b_{2g}e_{u}$ type, respectively) that easily explains an observed difference 
in numerical values and, even, in sign relations for  the corresponding copper and oxygen IC's 
\cite{Franck1996,Franck1993}.

For accounting of the pressure effect on $T_c$ through the $K_{eh}$ one need 
to determine full variation of the ground vibronic state of the PJT 
center including the pressure induced electronic density redistribution.
The pressure effect for the polar $CuO_4$-center can be described
by an effective operator
\begin{equation}
\hat{U}_p = \sum_i p_i^{(m)}({\bf p}) \hat{Q}_i +  \sum_i p_i^{(e)}({\bf p}) \hat{V}_i,
\end{equation}
where the first term describes the purely mechanical perturbation, and the second 
one corresponds to a pressure induced low-symmetry crystalline field, $Q_i$ and $\hat V_i$ 
are the distortion modes and electron operators with a certain symmetry, respectively. 
To get a linear in pressure variations $\delta \varepsilon_0$ and $\delta \varphi_0$ 
to the vibronic ground state energy $\varepsilon_0$ and wave function $\varphi_0$ 
in a framework of the Opik-Price method, one needs to solve the eigenvalue problem 
for the potential energy matrix
\begin{eqnarray}
&&\hat{U}(Q) = \hat{U}_0(Q) + \delta \hat{U}_p \ , \\
&&\hat{U}_0(Q) = \hat{H}_{el} + \sum_i \frac{\omega_i^2 Q_i^2}{2}\hat{I} + 
               \sum_i \hat{V}_i Q_i \ , \\ 
&&\delta \hat{U}_p = \sum_i \delta p_i^{(m)}Q_i \hat{I} +  \sum_i \delta p_i^{(e)}\hat{V}_i,
\end{eqnarray}
in the AP minima  with taking account of the linear in pressure variation of their coordinates:
\begin{equation}
Q_i = Q_i^{(0)} + \delta Q_i \ , \quad
\delta Q_i = - \frac{\delta p_i^{(n)}}{\omega_i^2} - 
 \frac{2 \langle \varphi_0|\hat{V}_i|\delta \varphi_0 \rangle}{\omega_i^2} . \label{Q}
\end{equation}
The system for $\delta \varepsilon_0$ and $\delta \varphi_0$  takes a form
\begin{eqnarray}
&& (\hat{H}_{el} + \sum_i \hat{V}_i Q_i^{(0)})|\delta\varphi_0 \rangle + \sum_i \hat{V}_i \left( \delta Q_i + \delta p_i^{(e)} \right)|\varphi_0 \rangle = 
\lambda_0 |\delta\varphi_0 \rangle + \delta \lambda_0 |\varphi_0 \rangle \nonumber \\
&& \langle \varphi_0 | \delta\varphi_0 \rangle = 0 , \quad
\delta \varepsilon_0 = \delta \lambda_0 - 
2 \sum_i \langle \varphi_0 |\hat{V}_i| \delta\varphi_0 \rangle Q_i^{(0)} . \label{sys} 
\end{eqnarray}

In addition,  pressure results in a variation of the energy surface form near 
a certain minimum of the adiabatic potential 
\[
\delta \epsilon = \sum_i
 \frac{2 \langle \varphi_0|\hat{U_1}| \varphi_i \rangle}{\lambda_0-\lambda_i}
 \left[ \langle \delta \varphi_0|\hat{U_1}| \varphi_i \rangle + 
        \langle \varphi_0|\hat{U_1}| \delta \varphi_i \rangle \right] 
\]
\begin{equation}
 - \sum_i \frac{\langle \varphi_0|\hat{U_1}| \varphi_i \rangle^2}{(\lambda_0-\lambda_i)^2}
(\delta\lambda_0-\delta\lambda_i) \label{eps},
\end{equation}
where $\hat{U}_1$ is the first variation of the vibronic Hamiltonian relative to $Q_i$; $\varphi_i$, $\lambda_i$ and $\delta \varphi_i$, $\delta \lambda_i$ are the eigenvectors and eigenvalues of the non-perturbed Opik-Price system and their variations under pressure, respectively. The system for $\delta \varphi_i$, $\delta \lambda_i$ is quite 
similar to (\ref{sys}). The second term in (\ref{eps}) results 
in a renormalization of the local frequencies, 
while the first one provides a hybridization of all the local modes.
The relations (\ref{Q})-(\ref{eps}) together with (\ref{K}) provide an algorithm 
for a calculation of the pressure dependence of the vibronic reduction factors.

In framework of the developed approach we have performed a number of model calculations of the 
baric coefficients with reasonable values of various parameters. Their results will be 
published elsewhere. As a whole, the PJT nature of 
the system explains both the main peculiarities and subtleties of the  pressure 
effects including  a strong anisotropy of  the baric coefficients in 123 system. 

At once, a real situation with isotope and baric effects in doped cuprates should account for 
phase separation  and percolation phenomena.


\section*{References}
\begin{thebibliography}{99}
\bibitem{Moskvin}  A.S. Moskvin, N.N. Loshkareva, Yu.P. Sukhorukov, 
M.A.Sidorov, A.A. Samokhvalov, \Journal{Sov. Phys. JETP}{78}{518}{1994}.

\bibitem{Alexandrov}  A.S. Alexandrov, \Journal{Phys. Rev. B}{46}{14932}{1992}.

\bibitem{MP}  A.S. Moskvin, Yu.D. Panov, \Journal{Sov. Phys. JETP}{84}{354}{1997)};
A.S. Moskvin, Yu.D. Panov, A.S. Ovchinnikov, M.A. Sidorov, \Journal{Physica C}{282-287}{1813}{1997}. 

\bibitem{Franck1996}  J.P. Franck, D.D. Lawrie, \Journal{J. Low Temp. Phys.}{105}{801}{1996}.

\bibitem{Franck1993}  J.P. Franck, S. Harker, J.H. Brewer, \Journal{Phys. Rev. Lett.}{71}{283}{1993}.

\end{thebibliography}
\end{document}